\documentclass{article}

\input{tcilatex}
\begin{document}

\title{Bell's Conjecture and Faster-Than-Light Communication}
\author{Luiz Carlos Ryff \\
\textit{Instituto de F\'{\i}sica, Universidade Federal do Rio de Janeiro,}\\
\textit{Caixa Postal 68528, 21941-972 RJ, Brazil}}
\maketitle

\begin{abstract}
The Bell-wave (B-wave) supposition has been introduced in an attempt to
investigate Bell's conjecture (according to which \textquotedblleft behind
the scenes something is going faster than \ light\textquotedblright ). Here
it is shown, for the case of two entangled photons, that if it is further
assumed that the B-waves propagate with superluminal but finite velocity
then it is possible, at least in principle, to have faster-than-light (FTL)
communication.
\end{abstract}

To try to explain the Einstein-Podolsky-Rosen (EPR) correlations, John Bell
conjectured that something should be propagating with superluminal velocity,
and suggested the reintroduction of the idea of an aether, a preferred frame
of reference \textrm{[1]}. However, as far as I know, he never elaborated on
this idea. B-waves have been assumed in an attempt to investigate Bell's
conjecture \textrm{[2]}. Considering a two-photon entangled state, a B-wave
is created when the first photon of the pair is detected in the preferred
frame. It then propagates and reaches the second photon, forcing it into a
well-defined state. But the state in which the first photon is found is not
necessarily the state into which the second photon will be forced. It will
depend not only on the initial entangled state, but also on the optical
devices the photons will find on their way to the detectors. How is the
correct information conveyed? Assuming that there cannot be any sort of
\textquotedblleft conspiracy\textquotedblright\ of nature, or, in other
words, that nature is \textquotedblleft blind\textquotedblright , this can
only take place in a purely mechanical or automatic way, so to speak. A
possibility is to have the B-wave following the path of the first photon
backwards to the source and then following the path of the second photon.
Each time it passes through an optical device its state is changed,
eventually reaching the second photon in the \textquotedblleft
correct\textquotedblright\ state. This simple mechanism can, in principle,
reproduce the results of all Bell inequalities tests with pairs of photons,
and is consistent with the following aspect of the quantum mechanical
formalism. For instance, let us consider the two-photon
polarization-entangled state $\mid \psi \rangle $ = (1/$\sqrt{2}$)($\mid $V$%
\rangle _{1}\mid $H$\rangle _{2}-\mid $H$\rangle _{1}\mid $V$\rangle _{2}$) $%
\equiv $ (1/$\sqrt{2}$)($\mid $V$\rangle \mid $H$\rangle -\mid $H$\rangle
\mid $V$\rangle $). If photon 1 passes through a $\lambda /2$-plate, such
that $\mid $V$\rangle $($\mid $H$\rangle $) $\rightarrow $ $\mid $H$\rangle $%
($\mid $V$\rangle $), then $\mid \psi \rangle $ $\rightarrow $ (1/$\sqrt{2}$%
)($\mid $H$\rangle \mid $H$\rangle -\mid $V$\rangle \mid $V$\rangle $). That
is, only the \textquotedblleft potential\textquotedblright\ states of photon
1 are changed (this is valid for any optical device, not only for wave
plates). It seems that the \textquotedblleft
communication\textquotedblright\ between the photons -- if it exists -- must
only occur at the moment of detection.

It has already been shown that the assumption of finite-speed ($v$)
superluminal communication leads, under specific circumstances and for more
than two entangled particles, to FTL signalling \textrm{[3]}. A simple
example is the following. Let us consider the state of three qubits $\mid $%
GHZ$\rangle $ = (1/$\sqrt{2}$)($\mid $0$\rangle _{1}\mid $0$\rangle _{2}\mid 
$0$\rangle _{3}$ +$\mid $1$\rangle _{1}\mid $1$\rangle _{2}\mid $1$\rangle
_{3}$). Particle 1 is sent to Alice (\textbf{A}) and particles 2 and 3 are
sent to Bob (\textbf{B}) and Charlie (\textbf{C}), respectively, who work
not far away from each other in the same lab.. At instant $t_{A}$ (in the
preferred frame), \textbf{A} may decide to measure the state, $\mid $0$%
\rangle $ or $\mid $1$\rangle $, of particle 1, or not; and, at instant $%
t_{L}>t_{A}$ (also in the preferred frame), \textbf{B} and \textbf{C} will
measure the state, $\mid $0$\rangle $ or $\mid $1$\rangle $, of their
particles. The condition $v>l/(t_{L}-t_{A})>c$ has to be fulfilled, where $l$
is the distance from \textbf{A }to\textbf{\ B } and from \textbf{A }to 
\textbf{C}. Supposing that the correlations are purely nonlocal, whenever 
\textbf{B} and \textbf{C} perform their measurements, but \textbf{A} does
not perform hers, the probability of \textbf{B} and \textbf{C} observing the
same outcome is $1/2$, since there can be no communication between them ($%
v<\infty $). On the other hand, whenever \textbf{B} and \textbf{C} perform
their measurements, and \textbf{A} performs hers, this probability is equal
to $1$, since the first measurement forces the other two particles into the
same state. Therefore, if we have in the left lab. many \textbf{A}s, and in
the right distant lab. the corresponding \textbf{B}s and \textbf{C}s, and
the \textbf{A}s combine to take the same decision together, that is, to
perform a measurement or not, the \textbf{B}s and \textbf{C}s will know
(comparing their results and disregarding improbable statistical
fluctuations) what has been decided in the left lab. before this information
could reach them transmitted by a light signal.

\bigskip For two particles, and assuming the existence of B-waves with the
properties mentioned above, the demonstration is as follows. Let us imagine
the following experiment performed in the preferred frame. A source $S$
emits entangled photons, $\nu _{1}$ and $\nu _{2}$, in state \textrm{[4]}%
\begin{equation}
\frac{1}{\sqrt{2}}\left( \mid V\rangle \mid H\rangle -\mid H\rangle \mid
V\rangle \right) .  \tag{1}
\end{equation}%
$\nu _{1}$ and $\nu _{2}$ are emitted in opposite directions, reaching
two-channel polarizers with orientations $\mathbf{a}$\textbf{\ }and $\mathbf{%
b}$\textbf{, }respectively. The condition 
\begin{equation}
x_{b}>x_{a}+2y  \tag{2}
\end{equation}%
is fulfilled, where $x_{b}$ is the distance followed by $\nu _{2}$ from $S$
to detector $D_{2}$($D_{2^{\prime }}$), placed on the transmission
(reflection) channel, and $x_{a}+2y$ is the distance followed by $\nu _{1}$
from $S$ to detector$\ D_{1}$($D_{1^{\prime }}$), placed on the transmission
(reflection) channel, and where $y$ is the height of a detour introduced in $%
\nu _{1}$'s path. Therefore, $\nu _{1}$ is always detected before $\nu _{2}$%
. Between $S$ and the detour there is a Pockels cell. Then, introducing $%
t_{l}$ and $t_{B}$ as 
\begin{equation}
t_{l}=\frac{x}{c}  \tag{3}
\end{equation}%
and 
\begin{equation}
t_{B}=\frac{x+2y}{v_{B}},  \tag{4}
\end{equation}%
where $x$ is the distance from the Pockels cell ($PC$) to $D_{1}$, and $%
v_{B} $ is the velocity of the Bell-wave, the condition 
\begin{equation}
t_{l}<t_{B}  \tag{5}
\end{equation}%
has to be fulfilled, which leads, using $(3)$ and $(4)$, to 
\begin{equation}
y>\left( \frac{v_{B}-c}{c}\right) \frac{x}{2}.  \tag{6}
\end{equation}%
Therefore, it is possible to have the detection of $\nu _{1}$ triggering a
light signal that activates the $PC$ just before the passage of the B-wave,
which, as a result, has its state modified. Since $v_{B}>c$, the B-wave
reaches $\nu _{2}$ before a light wave sent from $D_{1}$ or $D_{1^{\prime }}$
at the moment of detection. Now, let us assume that the activating signal is
only triggered when $\nu _{1}$ is registered at $D_{1^{\prime }}$. The
detection probabilities are then given by 
\begin{equation}
p_{12}=\frac{1}{2}\sin ^{2}(a,b),\text{ \ \ \ \ \ \ \ \ \ \ \ }p_{12^{\prime
}}=\frac{1}{2}\cos ^{2}(a,b),  \tag{7}
\end{equation}%
\begin{equation}
\text{ \ \ \ \ \ \ \ }p_{1^{\prime }2}=\frac{1}{2}\cos ^{2}(a,b^{\prime }),%
\text{ and \ \ \ }p_{1^{\prime }2^{\prime }}=\frac{1}{2}\sin
^{2}(a,b^{\prime }),\text{\ \ \ \ \ \ \ \ }  \tag{8}
\end{equation}%
where $\mathbf{b}^{\prime }\neq \mathbf{b}$, since the state of the B-wave
has been modified in accordance with our purposes. Hence, we obtain 
\begin{equation}
p_{12}+p_{12^{\prime }}=p_{1}=\frac{1}{2},  \tag{9}
\end{equation}%
\begin{equation}
p_{1^{\prime }2}+p_{1^{\prime }2^{\prime }}=p_{1^{\prime }}=\frac{1}{2}, 
\tag{10}
\end{equation}%
\begin{equation}
p_{12}+p_{1^{\prime }2}=p_{2}=\frac{1}{2}\left[ \sin ^{2}\left( a,b\right)
+\cos ^{2}\left( a,b^{\prime }\right) \right] ,  \tag{11}
\end{equation}%
and 
\begin{equation}
p_{12^{\prime }}+p_{1^{\prime }2^{\prime }}=p_{2^{\prime }}=\frac{1}{2}\left[
\cos ^{2}\left( a,b\right) +\sin ^{2}\left( a,b^{\prime }\right) \right] . 
\tag{12}
\end{equation}%
If the activating signal is not triggered, $\mathbf{b}^{\prime }\rightarrow 
\mathbf{b}$, which leads to $p_{2}=p_{2^{\prime }}=1/2$; on the other hand,
if it is, we have $p_{2}\neq p_{2^{\prime }}$. Therefore, comparing the
detections on the left side of the experimental apparatus it is possible to
know what decision was taken on the right side (to trigger the activating
signal or not); and this information can be transmitted with superluminal
velocity \textrm{[5]}.

The above discussion might be seen as an argument against the Bell
conjecture. However, if a preferred frame is assumed, the possibility of FTL
signaling can not be discarded; in particular, no causal paradoxes will
necessarily arise from this \textrm{[6]}. But some obstacles would make the
realization of the experiment difficult. In particular, we don't know how to
determine the preferred frame \textrm{[7] }and the superluminal speed. Using
recent data \textrm{[8]}, $(6)$ leads to $y>10^{4}x/2$, which suggests the
use of optical fibers, to keep the detour within the dimensions of the
laboratory!

Actually, it can always be conjectured that the communication between the
entangled photons does not occur through ordinary three-dimensional space;
however, this should not be an impediment\textrm{\ }to the investigation of
simple -- perhaps far too naive\textrm{\ }-- and experimentally testable
alternatives \textrm{[2]}, as the one discussed here.

\end{document}